\documentclass[aps,prd,notitlepage,showpacs,nofootinbib,superscriptaddress]{revtex4-1}

\usepackage{graphicx}
\usepackage[utf8]{inputenc} 

\usepackage{amsmath}
\usepackage{amssymb}
\usepackage{float}
\usepackage{comment}
\usepackage{slashed}
\usepackage[normalem]{ulem}

\usepackage{booktabs}
\usepackage{mathtools,braket,blkarray}

\usepackage[usenames,dvipsnames]{color}
\usepackage[colorinlistoftodos]{todonotes}
\usepackage[colorlinks=true,citecolor=darkred,urlcolor=darkred, pdfborder={0 0 0}]{hyperref}
\usepackage[normalem]{ulem}

\usepackage{multirow}
\definecolor{darkred}{rgb}{0.6,0,0}
\definecolor{brown}{rgb}{0.59, 0.29, 0.0}

\usepackage[colorinlistoftodos]{todonotes}

\definecolor{linkcolor}{rgb}{0,0,0.5}
 
 \newcommand {\ignore}[1]{}

\def \znbb {$\rm 0\nu\beta\beta$ }

\def\gsim{\raise0.3ex\hbox{$\;>$\kern-0.75em\raise-1.1ex\hbox{$\sim\;$}}}
\def\lsim{\raise0.3ex\hbox{$\;<$\kern-0.75em\raise-1.1ex\hbox{$\sim\;$}}}

\def\lnv{lepton number violating }
\def\SM{$\mathrm{SU(3)_c \otimes SU(2)_L \otimes U(1)_Y}$ }

\usepackage{soul}

\definecolor{mightnightblue}{RGB}{25,25,112}

\definecolor{brown}{rgb}{0.59, 0.29, 0.0}

\def\lnv{lepton number violation }

\def\vev#1{\left\langle #1\right\rangle}
\def\SM{$\mathrm{SU(3)_c \otimes SU(2)_L \otimes U(1)_Y}$ }
\def\21{$\mathrm{SU(2)_L \otimes U(1)_Y}$}

\def\lnv{lepton number violation }

\newcommand{\AddrAHEP}{%
  AHEP Group, Institut de F\'{i}sica Corpuscular --
  C.S.I.C./Universitat de Val\`{e}ncia, Parc Cient\'ific de Paterna.\\
  C/ Catedr\'atico Jos\'e Beltr\'an, 2 E-46980 Paterna (Valencia) - SPAIN}

\begin{document}
\bibliographystyle{unsrt}

\title{\boldmath \color{BrickRed}  Revamping Kaluza-Klein dark matter in an orbifold theory of flavor}

\author{Francisco J. de Anda}\email{fran@tepaits.mx}
\affiliation{Tepatitl{\'a}n's Institute for Theoretical Studies, C.P. 47600, Jalisco, M{\'e}xico}
\affiliation{Dual CP Institute of High Energy Physics, C.P. 28045, Colima, M\'exico}
\author{Omar Medina}\email{Omar.Medina@ific.uv.es}
\affiliation{\AddrAHEP}
\author{Jos\'{e} W. F. Valle}\email{valle@ific.uv.es}
\affiliation{\AddrAHEP}

\author{Carlos A. Vaquera-Araujo}\email{vaquera@fisica.ugto.mx}
\affiliation{Consejo Nacional de Ciencia y Tecnolog\'ia, Av. Insurgentes Sur 1582. Colonia Cr\'edito Constructor, Del. Benito Ju\'arez, C.P. 03940, Ciudad de M\'exico, M\'exico}
\affiliation{Departamento de F\'isica, DCI, Campus Le\'on, Universidad de
  Guanajuato, Loma del Bosque 103, Lomas del Campestre C.P. 37150, Le\'on, Guanajuato, M\'exico}
\affiliation{Dual CP Institute of High Energy Physics, C.P. 28045, Colima, M\'exico}


\begin{abstract}
\vspace{0.5cm}

We suggest a common origin for dark matter, neutrino mass and family symmetry within the orbifold theory proposed in~\cite{deAnda:2019jxw,deAnda:2020pti}.
Flavor physics is described by an $A_4$ family symmetry that results naturally from compactification.
WIMP Dark matter emerges from the first Kaluza-Klein excitation of the same scalar that drives family symmetry breaking and neutrino masses through the inverse seesaw mechanism.
In addition to the ``golden'' quark-lepton mass relation and neutrino predictions for \znbb decay, the model provides a good global description of all flavor observables.

\end{abstract}

\maketitle
\noindent

 \section{Introduction}

 Understanding flavor constitutes one of the major challenges in particle physics.
 Accounting for the observed pattern of fermion masses and the structure of their mixing parameters continues to defy the particle physics community,
even more so after the discovery of neutrino oscillations~\cite{McDonald:2016ixn,Kajita:2016cak,deSalas:2020pgw}.
The latter has demonstrated that leptons mix rather differently from the way quarks do in the Cabibbo–Kobayashi–Maskawa (CKM) model.
The Standard Model lacks a symmetry-based organizing principle that one may use to describe flavor properties. 
%
%
The striking oddness of the fermion pattern is unlikely to result just from randomness.
Rather, it indicates the presence of some ``family'' or ``flavor'' symmetry, a compilation of possibilities for discrete non-Abelian groups as family symmetry candidates is given in~\cite{Ishimori:2010au}. \\[-.4cm] 

The existence of extra space-time dimensions may shed light on the two complementary aspects of the flavor problem.
Indeed, while mass hierarchies may result from geometry~\cite{ArkaniHamed:1999dc}, 
mixing angle relations may be predicted from adequate symmetries~\cite{Chen:2015jta,Chen:2020udk},
thereby covering both sides of the coin.\\[-.4cm] 

Here we focus on six-dimensional orbifold theories proposed in~\cite{deAnda:2019jxw,deAnda:2020pti}. 
In contrast to the warped flavordynamics picture suggested in~\cite{Chen:2015jta,Chen:2020udk}, where family symmetries were imposed by hand,
here an $A_4$ family symmetry results naturally from compactification, a setup that was first introduced and described in detail in references \cite{Altarelli:2005yp,Altarelli:2006kg,Altarelli:2010gt}. 
Dark matter emerges as the first Kaluza-Klein (KK) excitation of the same scalar that drives lepton number violation, and family symmetry breakdown, leading to neutrino masses through an inverse seesaw mechanism. 
This new seed scalar field allows us to resurrect the KK dark matter proposal. The model leads naturally to the ``golden'' quark-lepton mass relation,
that implies strong predictions for the light-quark masses $m_d$ and $m_s$ and neutrinoless double beta decay.
  It also provides an adequate description of neutrino oscillation parameters, together with a very good global description of flavor observables. \\[-.4cm] 

This letter is structured as follows. In Sec.~\ref{sec:TheFrame} we recapitulate the theory framework of our model, while in Sec.~\ref{sec:kaluza-klein-wimp} we describe the general features of Kaluza-Klein dark matter,
followed in Sec.~\ref{sec:sigmakk-as-dm} by an analysis of the phenomenology of $\sigma^{KK}$ as a WIMP dark matter candidate.
In Sec.~\ref{app:FF} we present the ``golden quark-lepton'' mass formula, results for the neutrino sector, and also discuss our global fit to flavor observables.
In Sec.~\ref{sec:discussion-outlook-} we present a brief summary and additional discussion of our model.

\section{Theory Framework} 
\label{sec:TheFrame}

Our model is a 6-dimensional version of the Standard Model implementing the inverse seesaw mechanism~\cite{Mohapatra:1986bd,GonzalezGarcia:1988rw}.
It provides is a simple extension of the model presented in~\cite{deAnda:2019jxw}, where the orbifold compactification implies a discrete $A_4$ family symmetry in four dimensions \cite{Altarelli:2005yp,Altarelli:2006kg,Altarelli:2010gt}.  
  We stress that the 4-dimensional flavor symmetry is not arbitrary, but rather dictated by the extra-dimensional symmetries of our construction. 
  We assume a sequential setup where 3 singlet fermions $S$ accompany the 3 ``right-handed'' neutrinos. 
  The transformation properties of the fields under the gauge and family symmetry and their localization on the orbifold are shown in Table \ref{tab:fields}.
\begin{table}[h]
\centering
\footnotesize 
\begin{tabular}{ |cc|ccc|ccc|cc|c|}
\hline
\textbf{Field} &$\qquad$ & $SU(3)_C$ & $SU(2)_L$ & $U(1)_Y$ &$\qquad$  &$A_4$ &$\qquad$& $\mathbb{Z}_2$& $\mathbb{Z}_3$& Localization \\
\hline
$L$ & & $\mathbf{1}$ & $\mathbf{2}$ & $-1/2$ & & $\mathbf{3}$  & & $1$& $1$ &Brane\\
$d^c$ && $\bar{\mathbf{3}}$ & $\mathbf{1}$ & $1/3$ & & $\mathbf{3}$  & & $1$& $1$& Brane\\
$e^c$ &&$\mathbf{1}$ & $\mathbf{1}$ & $1$ & & $\mathbf{3}$  & & $1$& $1$& Brane\\
$Q$ && $\mathbf{3}$ & $\mathbf{2}$ & $1/6$ & & $\mathbf{3}$  & & $1$& $1$&Brane\\
$u_{1,2,3}^c$ && $\bar{\mathbf{3}}$ & $\mathbf{1}$ & $-2/3$ & & $\mathbf{1''},\mathbf{1'},\mathbf{1}$ & & $-1$& $1$&Bulk \\
$\nu^c$ && $\mathbf{1}$ & $\mathbf{1}$ & $0$ & & $\mathbf{3}$   & & $-1$&$\omega$&Brane \\
$S$ && $\mathbf{1}$ & $\mathbf{1}$ & $0$ & & $\mathbf{3}$   & & $-1$&$\omega^2$&Brane \\
\hline
$H_u$ && $\mathbf{1}$ & $\mathbf{2}$ & $1/2$ & & $\mathbf{3}$ & & $-1$&$1$& Brane \\
$H_d$  && $\mathbf{1}$ & $\mathbf{2}$ & $-1/2$ & & $\mathbf{3}$  & & $1$&$1$&Brane\\
$H_\nu$ && $\mathbf{1}$ & $\mathbf{2}$ & $1/2$ & & $\mathbf{3}$ & & $-1$&$\omega^2$& Brane \\
$\sigma$ && $\mathbf{1}$ & $\mathbf{1}$ & $0$ & & $\mathbf{3}$ & & $1$&$\omega^2$& Bulk\\
\hline
\end{tabular}
\caption{Field content of the model.}
\label{tab:fields}
\end{table}

The scalar sector consists of 3 Higgs doublets $H_u$, $H_d$ and $H_\nu$ plus an extra singlet scalar $\sigma$, all transforming as flavor triplets.  
Given the auxiliary discrete symmetries $\mathbb{Z}_2$ and $\mathbb{Z}_3$ in Table \ref{tab:fields}, $H_d$ only couples to down type fermions (charged leptons and down quarks),
while $H_u$ couples only to up quarks and $H_\nu$ only couples to neutrinos. 
The effective Yukawa terms are given by 
\begin{equation}
\begin{split}
\mathcal{L}_Y &= y_1^\nu(L H_\nu \nu^c)_1+y_2^\nu(L H_\nu \nu^c)_2+\frac{y^S}{2}\sigma S S+ m_{\nu^c S}\nu^c S\\
 &\quad +y_1^d(Q d^c H_d)_1+y_2^d(Q d^c H_d)_2+y_1^e(L e^c H_d)_1+y_2^e(L e^c H_d)_2\\
&\quad +y_1^u(QH_u)_{1'}u_{1}^c+y_2^u(QH_u)_{1''}u_{2}^c+y_3^u(QH_u)_{1}u_{3}^c,
\label{eq:yuk}
\end{split}
\end{equation}
where $()_{1,2}$ and $()_{1,1',1''}$ indicate possible singlet contractions $\mathbf{3}\times \mathbf{3} \times \mathbf{3}\to \mathbf{1}_{1,2}$ and $\mathbf{3}\times \mathbf{3} \to \mathbf{1}_{1,1',1''}$ in $A_4$.
Here we will also assume all dimensionless Yukawa couplings to be real. \\[-.4cm] 

Moreover, we adopt a dynamical scenario where lepton number and family symmetries are violated spontaneously, through the expectation value of an extra scalar field $\sigma$,
whose vacuum expectation value (VEV) yields Majorana masses to the singlet fermion $S$. 
The VEV of $\sigma$ must lie in the zero mode of the extra dimensional field decomposition ~\cite{Kobayashi:2008ih}, and must comply with the $\mathbb{Z}_2$ boundary condition 
\begin{equation}
P\braket{\sigma}=\braket{\sigma}.
\end{equation}
Here $P$ defines an arbitrary gauge twist of the orbifold, associated to its geometry, and is chosen to be  
\begin{equation}
P=\frac{1}{3}\left(\begin{array}{ccc}
-1 & 2\omega^2 & 2\omega\\
2\omega & -1 & 2\omega^2 \\
2\omega^2 & 2\omega & -1
\end{array}\right),
\label{eq:boun}
\end{equation}
with $\omega=e^{2\pi i/3}$, the cube root of unity. Notice that it satisfies $P^2=1$. This boundary condition aligns the VEV as 
\begin{equation}
\braket{\sigma}=v_\sigma\left(\begin{array}{c}1\\ \omega\\ \omega^2\end{array}\right),
\end{equation}

The Higgs doublets are assumed to obtain the most general vacuum alignment, which we parametrize as in \cite{deAnda:2019jxw}, these alignments require in general $A_4$ soft breaking terms in the scalar potential \cite{Morisi:2011pt}, nonetheless they are a possibility for obtaining viable neutrino mixing results in models with three Higgs-doublets as an $A_4$ triplet \cite{GonzalezFelipe:2013yhh},%
\begin{equation}\begin{split}
\braket{H_u}=v_u\left(\begin{array}{c}\epsilon_1^u e^{i\phi_1^u}\\ \epsilon_2^u e^{i\phi_2^u}\\ 1\end{array}\right),\ \ \ \ \braket{H_\nu}=v_\nu e^{i\phi^\nu}\left(\begin{array}{c}\epsilon_1^\nu e^{i\phi_1^\nu}\\ \epsilon_2^\nu e^{i\phi_2^\nu}\\ 1\end{array}\right),\ \ \ \braket{H_d}=v_d e^{i\phi^d}\left(\begin{array}{c}\epsilon_1^d e^{i\phi_1^d}\\ \epsilon_2^d e^{i\phi_2^d}\\ 1\end{array}\right).
\end{split}\end{equation}
The explicit form of the mass matrices for the quarks and charged leptons (up to unphysical rephasings) is given as
\begin{gather}
 M_u=v_u\left(\begin{array}{ccc} y_1^u\epsilon_1^u &y_2^u \epsilon_1^u & y_3^u\epsilon_1^u \\
 y_1^u\epsilon_2^u \omega^2&  y_2^u\epsilon_2^u \omega &   y_3^u\epsilon^u_2 \\
  y_1^u \omega& y_2^u \omega^2&y_3^u\end{array}\right),\ \ \ 
M_d=v_d\left(\begin{array}{ccc} 0 & y_1^d\epsilon_1^d e^{i(\phi_1^d-\phi_2^d)}& y_2^d \epsilon_2^d \\
 y_2^d\epsilon_1^d e^{i(\phi_1^d-\phi_2^d)}& 0 &  y_1^d\\
 y_1^d\epsilon_2^d & y_2^d&0\end{array}\right),\nonumber\\
M_e=v_d\left(\begin{array}{ccc} 0 & y_1^e\epsilon_1^d e^{-i(\phi_1^d-\phi_2^d)}& y_2^e \epsilon_2^d \\
 y_2^e\epsilon_1^d e^{-i(\phi^1_d-\phi_2^d)}& 0 &  y_1^e\\
 y_1^e\epsilon_2^d & y_2^e&0\end{array}\right),\label{eq:massmat1}
\end{gather}
Turning now to the neutral fermion mass matrix, in the $(\nu,\nu^c, S)$ basis, it is given as 
\begin{equation}
M_{\nu}=\left(\begin{array}{ccc}
0& M_D &0\\
M^T_D&0& M_{\nu^c S}\\
0 & M^T_{\nu^c S}& M_S
\end{array}\right),
\end{equation}
with
\begin{gather}
  M_D=v_\nu\left(\begin{array}{ccc} 0 & y_1^\nu\epsilon_1^\nu e^{i(\phi^\nu_1-\phi_2^\nu)} & y_2^\nu \epsilon_2^\nu \\
 y_2^\nu\epsilon_1^\nu e^{i(\phi_1^\nu-\phi_2^\nu)}& 0 &  y_1^\nu\\
 y_1^\nu\epsilon_2^\nu & y_2^\nu&0\end{array}\right),\nonumber\\
 M_{\nu^c S}=m_{\nu^c S}~\left(\begin{array}{ccc}
 1 & 0 & 0 \\
 0 & 1 & 0 \\  
 0 & 0 & 1
 \end{array}\right),\ \ \
M_S=y^S v_\sigma\left(\begin{array}{ccc}0 &   \omega^2 & \omega \\  \omega^2 & 0&1\\  \omega&1&0\end{array}\right).
\label{eq:massmat2}
\end{gather}

After spontaneous symmetry breaking, the light neutrinos acquire masses through the inverse seesaw mechanism, characterized by the effective mass matrix~\cite{Mohapatra:1986bd,GonzalezGarcia:1988rw}
\begin{equation}
m_\nu\sim M_D M^{-1}_{\nu^c S} M_S M^{T-1}_{\nu^c S}M_D^T=\frac{M_D M_S M_D^T}{m_{\nu^c S}^2}.
\end{equation}
Notice that, due to the family symmetry, the singlet mass entry $\nu^c S$ is trivial and the others, $M_D$ and $M_S$, have zeros along the diagonal.
We will assume the following hierarchy of scales 
\begin{equation}
M_D\sim \braket{H_\nu}\sim \mathcal{O}(\mathrm{GeV}), \ \ \ M_S\sim \braket{\sigma}\lsim \mathcal{O}(\mathrm{GeV}), \ \ m_{\nu^c S} \sim \mathcal{O}(10\, \mathrm{TeV}).
\label{eq:lsss}
\end{equation}
This choice renders naturally light neutrino masses.

\section{Revamping Kaluza-Klein Dark matter} 
\label{sec:kaluza-klein-wimp}

Besides the gauge fields, our model contains only two fields that propagate into the bulk, namely $u^c_i$ and $\sigma$, where the latter is an electrically neutral scalar transforming
as a singlet under the symmetries of the \SM gauge group. In this section we examine the possibility of identifying the lightest Kaluza-Klein mode of $\sigma$ as a WIMP dark matter candidate.

The extra dimensions are orbifolded by a $\mathbb{Z}_2$. Every field has a different $\mathbb{Z}_2$ charge, depending on its $A_4$ and Lorentz transformation.
In principle, one could choose the charge for each field individually, which implies that some couplings are also charged under the $\mathbb{Z}_2$ orbifolding, usually called kink couplings.
We instead assume that each coupling constant in the model is trivial under orbifolding.
This leaves the $\mathbb{Z}_2$ orbifolding symmetry as a symmetry of the compactified Lagrangian.
This surviving $\mathbb{Z}_2$ (not to be confused with the discrete auxiliary symmetry in Table \ref{tab:fields}) protects the lightest KK mode,
making it stable and a potential dark matter candidate. \\[-.4cm] 

In order to determine the mass scale of the lightest KK mode, we start by analyzing the spectrum of the effective 4-dimensional theory that emerges after the $T^2/\mathbb{Z}_2$
orbifold compactification of the extra two dimensions present in our construction. 
Denoting the extra dimensions by a single complex coordinate $z$, we can decompose any field in the bulk $\Psi$ (fermion, scalar or vector)
into a tower of low-energy 4-dimensional effective modes $\Psi_{ni}(x)$ as 
\begin{equation}
\Psi(x,z)=\sum_{n=0}^{\infty}\sum_{i=\pm}\Psi_{ni}(x)f_{ni}(z),
\label{eq:deco}
\end{equation}
where the profiles $f_{ni}(z)$ are eigenfunctions of the extra dimensional translation  and the $\mathbb{Z}_2$ orbifold parity, these satisfy by definition the following conditions
\begin{equation}
  f_{n\pm}(z)=f_{n\pm}(z+1)=f_{n\pm}(z+\omega)=\pm f_{n\pm}(-z).
  \label{eq:fcon}
\end{equation}
Note that the profiles form an orthonormal basis
\begin{equation}
\int dzd\bar{z} f^*_{ni}(z)f_{mj}(z)=\delta^m_n\delta^i_j.
\label{eq:fort}
\end{equation}
 As a result of the periodicity conditions in Eq.~(\ref{eq:fcon}) and orthonormality from Eq.~(\ref{eq:fort}), the profiles must be translation eigenfunctions
  (which for flat extra dimensions are exponentials) with
\begin{equation}
\partial_z f_{ni}=M_{ni} f_{ni},
\end{equation}
with  $M_{ni}$ proportional to $M_{ni}\sim n/R$ where $R$ is the compactification scale. 
Hence, for example, the kinetic term for a 6-dimensional scalar field becomes a collection of mass terms for their corresponding 4-dimensional effective degrees of freedom  
\begin{equation}
\int dz d\bar{z} (\partial_z\Psi)^\dagger \partial_z\Psi=\sum_{n,m,i,j}M_{ni}M_{mj}\Psi_{ni}^\dagger\Psi_{mj}\int dz d\bar{z} f_{ni}^* f_{mj}=\sum_{n,i}M_{ni}^2\Psi_{ni}^\dagger \Psi_{ni},
\end{equation}
with an analogous relation for fermions. By choosing an orthonormal basis, the KK modes are already in the mass basis.
Note that the zero-mode has an eigenvalue $M_{0i}=0$, so that its profile $f_{0i}$ is a constant.
The zero-modes remain massless after compactification, and can be identified at low energies with the SM fields. \\[-.4cm]

Concerning the heavy KK modes, there is an important result that comes from the preservation of the orbifolding $\mathbb{Z}_2$ symmetry. 
If one tries to build a term involving a single heavy KK mode (with $n\geq 1$) and any number $N$ of massless zero modes $\Psi_0^{(k)}$, $k=1,\dots,N$,
one finds that such term automatically vanishes, as the effective 4-dimensional coupling comes from integrating the extra dimensional profiles
\begin{equation}
\begin{split}
\int dz d\bar{z} \Psi_{0}^{(1)}\cdot...\cdot \Psi_0^{(N)}\Psi_{nj} f_0^N f_{nj}&=\Psi_{0}^{(1)}\cdot...\cdot \Psi_0^{(N)}\Psi_{nj} f_0^N\ \ \int dz d\bar{z}  f_0 f_{nj}\\
&=
\Psi_{0}^{(1)}\cdot...\cdot \Psi_0^{(N)}\Psi_{nj}f_0^{N-1} \delta^n_0=0.
\end{split}
\end{equation}
The first line follows from the fact that the $\Psi$ functions do not depend on the extra dimensions, and the zero mode profiles are just a constant.
The second line results from the orthonormality of the profiles. Such terms would in general induce the decay of KK modes into $N$ massless modes.
However, in our construction these terms clearly do not exist. There might be destabilizing interactions between KK modes.
However, only the ``right-handed'' up-type quarks $u^c$ and the $\sigma$ scalar have KK modes, and these do not interact with each other.
Therefore there are no destabilizing interactions and the lightest KK mode is stable.

In the present model there are three types of fields in the bulk, the gauge fields, the ``right-handed'' up-type quarks $u^c$ and the $\sigma$ scalar driving lepton number and
family symmetry breaking. 
 In the following we will assume that the lightest KK mode is electrically neutral. That leaves us with the first KK modes of the neutral SM gauge bosons:
 the photon, gluon or the $Z$ boson~\cite{Servant:2002aq}, as well as first KK mode of the $\sigma$ scalar.
 The latter is charged only under $A_4$ and drives family symmetry breaking and neutrino mass generation. 
 The lightest KK excitation of this field is a potential dark matter candidate. 
 After orbifold compactification and spontaneous electroweak symmetry breaking the first neutral KK modes will acquire masses of the order  
\begin{equation}
M_{(\gamma,g)^{KK}}^2\sim\left(\frac{1}{2R}\right)^2,\ \ \ M_{Z^{KK}}^2\sim\left(\frac{1}{2R}\right)^2+m_Z^2,\ \ \ M_{\sigma^{KK}}^2\sim\left(\frac{1}{2R}\right)^2+m_\sigma^2.
\label{eq:lkp}
\end{equation}

In the simplest extra dimensional extension of the SM it is natural to assume that the lightest KK particle (LKP) is the first mode of the photon or the gluon,
as their zero mode always remains massless. \\[-.4cm]

Note that Eq.(\ref{eq:lkp}) only gives the order of magnitude of the first excitations, any mass hierarchy can be assumed for the neutral KK light modes. 
If one assumes that the LKP is a dark matter candidate and identifies it with the first excitation of the photon, the gluon or the $Z$ gauge boson,
it is in general difficult to successfully reproduce the observed dark matter relic abundance, since the all gauge couplings are completely fixed, 
their masses being the only free parameter, determined by the compactifcation scale.
In particular, if the first KK mode of the photon is identified with the LKP, in order to correctly reproduce the required relic abundance 
a mass of order $M_{(\gamma,Z)^{KK}}\sim 1\,\mathrm{TeV} $ is required \cite{Burnell:2005hm,Arrenberg:2008wy}. 
Similarly, if the LKP is the first KK gluon then its mass must be of order $M_{g^{KK}}\sim 5\, \mathrm{TeV}$ \cite{Kong:2005hn}. 
Note that both scenarios predict compactification scales that lie in tension the LEP bound in Eq.~(\ref{eq:compac}). 
Furthermore, the case of vector dark matter is currently disfavored experimentally \cite{Dong:2017zxo}. 

Previous proposals~\cite{Servant:2002aq,Kong:2005hn,Burnell:2005hm,Arrenberg:2008wy} with the LKP as a dark matter candidate,
  assumed the LKP to be associated to a Standard Model field, a possibility no longer viable experimentally.
  In our present setup, the LKP comes from a genuinely new field, beyond those of the Standard Model.
  This field drives \lnv plays a key role in explaining the flavor structure of the fermions \cite{deAnda:2019jxw,deAnda:2020pti}.
  This choice is not only well-motivated but also relaxes the constraints, allowing us to build a viable and predictive model.
Indeed, we will show that if the first KK mode of $\sigma$ is identified with the LKP it can play the role of a WIMP dark matter candidate. 

Before closing this section, we comment on the key role played by the compactification scale on the Kaluza-Klein WIMP dark matter phenomenology.
It has long been noted that the existence of a KK tower generates flavor Changing Neutral Currents (FCNCs).
These arise when the extra-dimensional fields in the bulk are allowed to have an explicit mass term, which
prevents the simultaneous flavor diagonalization of their higher KK modes with the zeroth level Lagrangian \cite{Cheung:2001mq,Barbieri:2004qk}.  
We note, however, that our model is built in six flat dimensions, and the bulk fermions are 6-D chiral fields, with no explicit mass terms.  
As a result, it is free from FCNCs of this kind by construction.  
However, an important effect arises from the additional contributions to the Electroweak Precision observables.
The Peskin-Takeuchi $S$, $T$ and $U$ parameters are modified by the existence of a tower of massive vector $\mathrm{SU(2)_L}$ triplets.  
The current experimental bound for a setup with 2 flat non-Universal Extra Dimensions is~\cite{Deutschmann:2017bth,Ganguly:2018pzs}  
\begin{equation}
\frac{1}{2R} \gtrsim 2.1\ {\rm TeV}.
\label{eq:compac}
\end{equation}
As shown in Eq.(\ref{eq:lkp}), the mass of the first neutral KK mode for $\sigma$ has two contributions.
The first one comes from the compactification scale constrained above.  
The second contribution $m_\sigma^2$ is actually negative, since the zero-mode $\sigma$ must develop a VEV $\braket{\sigma}$ after spontaneous symmetry breaking.  
This triggers neutrino mass generation by the inverse seesaw mechanism~\cite{Mohapatra:1986bd,GonzalezGarcia:1988rw} and breaks the family symmetry as well.
Its associated scale $\braket{\sigma}$ is estimated in Eq. (\ref{eq:lsss}).  
As we will show in the next section the model can accommodate naturally a mass of a few TeV for the LKP, which can provide a viable WIMP dark matter particle.

\section{$\sigma^{KK}$ as a viable WIMP dark matter candidate}   
\label{sec:sigmakk-as-dm}

Here we show how the lightest KK mode of the field $\sigma$, identified as the LKP, can be a successful WIMP dark matter candidate,
while the corresponding scalar simultaneously provides small neutrino masses through a low-scale inverse seesaw mechanism~\cite{Mohapatra:1986bd,GonzalezGarcia:1988rw},
and the breakdown of a flavour symmetry following the lines proposed in~\cite{deAnda:2019jxw,deAnda:2020pti}.   \\[-.4cm]

In our model the complex scalar singlet $\sigma$ transforms as a triplet under the $A_4$ flavor symmetry, providing the source of $A_4$ flavor symmetry breaking,
driving low-scale neutrino mass generation through spontaneous violation of lepton number.
 We now discuss the role of the first KK mode of $\sigma$ as a DM candidate in our framework.
 In contrast to the case of the gauge bosons, its  couplings are not fixed by SM interactions.
 As a result they can be used to fit the observed relic abundance and direct detection constraints \cite{Kong:2005hn},  
 together with a viable compactification scale and a natural mechanism for light neutrino masses.\\[-.4cm]

{After compactification, one obtains a 4-dimensional scalar potential whose relevant terms can be written generically as 
\begin{equation}
\begin{split}
\mathcal{L}_\sigma=&\ \mu_\sigma^2\sigma^\dagger\sigma+\lambda_\sigma (\sigma^\dagger\sigma)^2+\tilde{\lambda}_\sigma (\sigma^\dagger\sigma\sigma^\dagger\sigma)+\lambda_{u,d,\nu}\sigma^\dagger\sigma H_{u,d,\nu}^\dagger H_{u,d,\nu}.
\label{eq:scpot}
\end{split}
\end{equation}
In the above relation, the $\lambda_\sigma$ term involves the  $3\times 3\to 1$  contraction under the $A_4$ symmetry, 
while the $\tilde{\lambda}_\sigma$ term characterizes the symmetric $3\times 3\times 3\times 3\to 1$ contraction. 
The coefficient $\mu^2_\sigma$ is negative, in consistency with the spontaneous symmetry breakdown driven by the zero mode of $\sigma$.
The corresponding VEV becomes 
\begin{equation}
v_\sigma^2=\frac{-\mu_\sigma^2}{2\lambda_\sigma+6\tilde{\lambda}_\sigma},
\end{equation}
where we have assumed that the alignment of $\sigma$ is fixed by the extra dimensional boundary condition, entering as an input to the potential. 
Note that the scale of $v_\sigma$ is determined by the arbitrary parameter $\mu^2_\sigma$ which is unrelated to the compactification scale. 
An important consequence of the spontaneous symmetry breaking is that Eq.(\ref{eq:lkp}) is rewritten as
\begin{equation}
M_{\sigma^{KK}}^2\sim\left(\frac{1}{2R}\right)^2+\mu^2_\sigma,
\label{eq:lkp2}
\end{equation}
with $\mu^2_\sigma<0$, which allows the first excited mode of $\sigma$ to be identified as the LKP and, as we will see, implement the WIMP dark matter picture.

We can further improve the estimate for the LKP mass by decomposing a generic bulk field into an infinite tower of orbifold eigenstates, as shown in Eq.(\ref{eq:deco}).  
  Each 6-dimensional field should be decomposed into a series of the extra dimensional translation and reflection eigenstates
For example the bulk $\sigma$ field, which is a flavor triplet, should be decomposed as a linear combination of the three eigenvectors of the boundary condition matrix in Eq. \ref{eq:boun}.
Therefore, its full decomposition into orbifold eigenstates is 
\begin{equation}
\sigma(x,z)=\sum_{n=0}^\infty\left[\sigma_{n+}(x)\frac{1}{\sqrt{3}}\left(\begin{array}{c}1\\ \omega\\ \omega^2\end{array}\right)f_{n+}(z)+\sigma_{n-}(x)\frac{1}{\sqrt{2}}\left(\begin{array}{c}1\\ 0\\ -\omega^2\end{array}\right)f_{n-}(z)+\tilde{\sigma}_{n-}(x)\frac{1}{\sqrt{2}}\left(\begin{array}{c}0\\ 1\\ -\omega\end{array}\right)f_{n-}(z)\right],
\end{equation}
where the three 4-dimensional towers of fields $\sigma_{n+},\sigma_{n-},\tilde{\sigma}_{n-}$ represent the components of the $A_4$ triplet. 
Here the labels $\pm$ denote the  eigenvectors corresponding to the $\pm 1$ eigenvalues of the extra dimensional boundary condition in Eq.(\ref{eq:boun}).  
In this notation, the zero mode is associated with a constant profile $f_{0+}$ and a vanishing $f_{0-}$.
The LKP would correspond to the first mode, which can be any of the three complex scalars 
\begin{equation}
\sigma_{1LKP}=\sigma_{1+}(x)\frac{1}{\sqrt{3}}\left(\begin{array}{c}1\\ \omega\\ \omega^2\end{array}\right),\ \ \ \sigma_{2LKP}=\sigma_{1-}(x)\frac{1}{\sqrt{2}}\left(\begin{array}{c}1\\ 0\\ -\omega^2\end{array}\right),\ \ \ \sigma_{3LKP}=\tilde{\sigma}_{1-}(x)\frac{1}{\sqrt{2}}\left(\begin{array}{c}0\\ 1\\ -\omega\end{array}\right).
\end{equation}
Due to the orbifolding, these obtain the same mass $\sim 1/(2R)$ after compactification. 
The degeneracy is lifted by the potential and the zero mode VEV $v_\sigma$. However, under the natural assumption $1/(2R)\gg v_\sigma$, the mass splitting is expected to be small.   

From the scalar potential in Eq.(\ref{eq:scpot}) we can obtain the mass contributions arising from $v_\sigma$ as 
\begin{equation}\begin{split}
\mathcal{L}_{m\sigma}=
\Big(\sigma_{1+}^*,\sigma_{1-}^*,\tilde{\sigma}_{1-}^*\Big)\left(\begin{array}{ccc}
\frac{1}{4R^2}-(5\lambda_\sigma+7\tilde{\lambda}_\sigma)v^2_\sigma & 0 & 0 \\
0& \frac{1}{4R^2}-\frac{10\lambda_\sigma+11\tilde{\lambda}_\sigma}{2}v^2_\sigma
& -\frac{5\omega^2\lambda_\sigma+(2-\omega-5\omega^2)\tilde{\lambda}_\sigma}{2}v^2_\sigma
\\ 0& -\frac{5\omega\lambda_\sigma+(2-\omega^2-5\omega)\tilde{\lambda}_\sigma}{2}v^2_\sigma & \frac{1}{4R^2}-\frac{10\lambda_\sigma+11\tilde{\lambda}_\sigma}{2}v^2_\sigma  
\end{array}\right)
\left(\begin{array}{c}
\sigma_{1+}\\ \sigma_{1-} \\ \tilde{\sigma}_{1-} \\
\end{array}\right).
\label{eq:sigmasfla}
\end{split}
\end{equation}

Barring the presence of new co-annihilation channels, the parameter space allowed by the observed relic complex-scalar dark matter abundance and direct detection is in general very constrained. 
In our model the parameter space is widened by the fact that the KK modes are nearly degenerate in mass, leading naturally to a multicomponent dark matter picture. 

In order to demonstrate the viability of our model as a theory for dark matter, it suffices to consider a simplified scenario in which all the non-SM fields are heavy and decouple,
except for the almost degenerate complex scalars $\sigma_{1+}$, $\sigma_{1-}$ and $\tilde{\sigma}_{1-}$. 
For illustration we will include only one Higgs doublet $H$, omitting the explicit $A_4$ contractions. 
This way the relevant Higgs portal parameters are the effective quartic coupling constants of the extended scalar sector characterizing the potential
\begin{equation}
\begin{split}
V(H,\sigma_{1-},\tilde{\sigma}_{1-})\supset&\lambda_1 H^\dagger H \sigma_{1+}^*\sigma_{1+}+\lambda_2 H^\dagger H \sigma_{1-}^*\sigma_{1-}+\lambda_3 H^\dagger H \tilde{\sigma}_{1-}^*\tilde{\sigma}_{1-}\\&+\lambda_4H^\dagger H(\sigma_{1+}^*\sigma_{1-}+\sigma_{1-}^*\sigma_{1+})+\lambda_5H^\dagger H(\sigma_{1+}^*\tilde{\sigma}_{1-}+\tilde{\sigma}_{1-}^*\sigma_{1+})+\lambda_6H^\dagger H(\tilde{\sigma}_{1-}^*\sigma_{1-}+\sigma_{1-}^*\tilde{\sigma}_{1-}).
\end{split}
\end{equation}
After electroweak spontaneous symmetry breaking, the complex scalars $\sigma_{1+}$,  $\sigma_{1-}$ and $\tilde{\sigma}_{1-}$ mix into neary degenerate physical complex scalars $\phi_1$, $\phi_2$ and $\phi_2$.  

In order to ensure perturbativity we have studied the relic abundance and direct detection constraints for this simplified scenario.
We have varied randomly the relevant parameters in the range $0<\lambda_{i}<\sqrt{4\pi}$.
Moreover, we have modeled a small mass splitting amongst the three physical scalars identifying $\phi_1$ as the LKP dark matter candidate,
and scanning the mass parameters within the range $0<m_{\phi_1}=m_{DM}< 10^4\,\mathrm{GeV}$, $0<m_{\phi_2},m_{\phi_3}< 1.2\times m_{DM}$. \\[-.4cm] 

 The results are shown in Fig.\ref{fig:DMplot}. Each magenta point corresponds to a model parameter combination that reproduces the correct relic abundance $\Omega h^2= 0.120$ \cite{Planck:2018vyg} 
 through Higgs portal interactions, and enhanced by re-scattering with the other nearly degenerate complex scalars in the first KK-mode sector.
 This way the model opens a rather large parameter window below the current direct detection constraints, including the recent ones of LUX-ZEPLIN \cite{LUX-ZEPLIN:2022qhg}.
 Many of these points will be probed within upcoming dark matter experiments. \cite{Billard:2021uyg,DarkSide-20k:2017zyg,Schumann:2015cpa} 

In conclusion we find that our LKP, the lightest complex scalar contained in the first excited mode of $\sigma$, can provide a viable WIMP dark matter candidate. 

\begin{figure}[H]
\begin{center}
\includegraphics[width=0.7\textwidth]{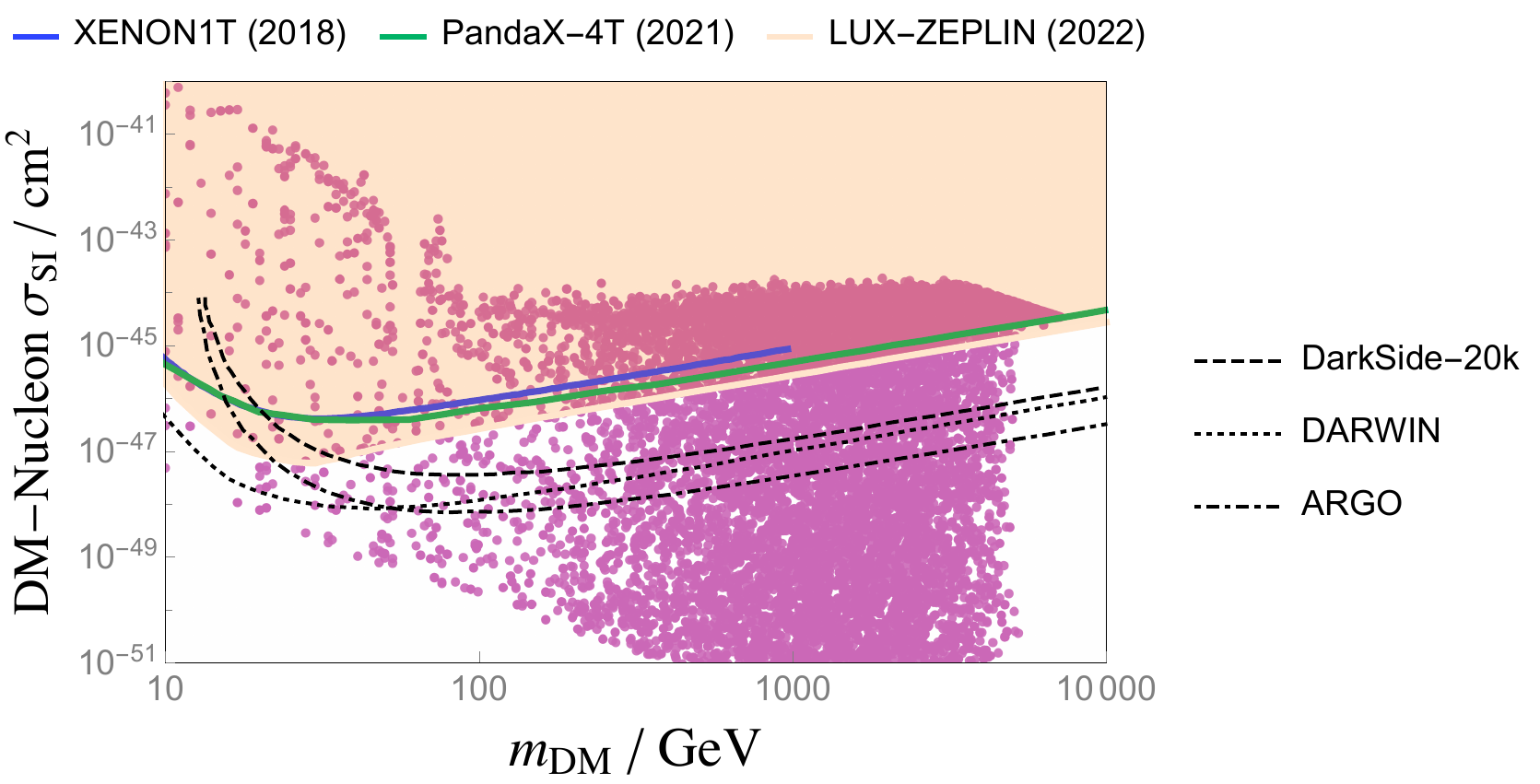}
\caption{ The direct detection and relic abundance constraints versus the dark matter mass $m_{DM}=m_{\sigma_{1-}}$.
  Each magenta point corresponds to a combination of the model parameters that give the correct relic abundance $\Omega h^2= 0.120$.
  The orange shaded region is ruled out by the LUX-ZEPLIN experiment \cite{LUX-ZEPLIN:2022qhg}. We also show the PandaX-4T \cite{PandaX-4T:2021bab} and XENON1T \cite{XENON:2018voc} limits,
  and the projected sensitivities of some upcoming direct detection experiments.\cite{Billard:2021uyg,DarkSide-20k:2017zyg,Schumann:2015cpa}}
\label{fig:DMplot}
\end{center}
\end{figure}
\section{Flavor predictions and global fit}
\label{app:FF}

Our model is based on an $A_4 \otimes \mathbb{Z}_2 \otimes \mathbb{Z}_3$ family symmetry as summarized in Table \ref{tab:fields}, with a total of 17 independent parameters that determine the flavour
properties of both the quark and lepton sectors. These parameters, described in equations (\ref{eq:yuk})-(\ref{eq:massmat2}), can be written compactly as  
\begin{equation}
 \left\{ \frac{(y_{1,2}^{\nu}v_\nu)^2 y^S v_{\sigma}}{m^2_{\nu^c S}},y_{1,2}^{e,d}v_d,\ y_{1,2,3}^uv_u,\ \epsilon_{1,2}^{u,\nu,d}, \ \ \phi_1^{\nu,d}-\phi_2^{\nu,d}\right\}.
 \label{eq:parameters}
\end{equation}

 We have performed a goodness of fit analysis of the model using a reduced chi-squared test. The reference values for the flavour related observables for the quark and lepton sectors
 were taken from references~\cite{deSalas:2020pgw,Antusch:2013jca,Huang:2020hdv}, and the masses of the SM fermions were taken at the Z-boson mass ($M_Z$) scale. 

 Our chi-square test assesses how well the model is able to describe the 19 experimentally determined low-energy flavor observables,
 including the SM fermion masses, the quark mixing parameters, and the neutrino oscillation parameters,
\begin{equation}
 \left\{m_{u,c,t,d,s,b,e,\mu,\tau},\Delta m^{2}_{21}, \Delta m^{2}_{31} ,\ \theta_{12,13,23}^q, \theta^{\ell}_{12,13,23}, \delta^q,\delta^{\ell}\right\},
 \label{eq:obs}
\end{equation}
therefore the number of degrees of freedom $K=19-17=2$. 

Table \ref{tab:fit2} summarizes the results from our goodness-of-fit global analysis of flavor observables.
  The values for the different charged lepton masses and mixing angles are taken from~\cite{Antusch:2013jca}.
  For the quark masses we used an updated determination from reference~\cite{Huang:2020hdv}.
For definiteness we assume the experimentally preferred case of normal ordered (NO) neutrino spectrum. 
Our best fit point, as described in detail by Table \ref{tab:fit2}, yields $\chi^{2}_{K} = \frac{\chi^2}{K}=1.05$,
showing that the model is in very good agreement with the experimental data. \\[-.4cm]

\subsection{ ``Golden" Quark-Lepton Mass Formula}
\label{sec:golden-quark-lepton}

A key prediction of our model comes from the fact that the charged leptons and down-quarks, both transforming as triplets of $A_4$, obtain their masses from the coupling with the same doublet Higgs, $H_d$. 
The flavor symmetry structure of the mass matrices $M_e$, and $M_d$ in equation (\ref{eq:massmat1}) implies the following relation between their masses 
\begin{equation}\label{eq:golden}
\frac{m_\tau}{\sqrt{m_\mu m_e}}=\frac{m_b}{\sqrt{m_s m_d}},
\end{equation}
This relation has been called ``golden quark-lepton mass formula''~\cite{Morisi:2011pt,King:2013hj,Morisi:2013eca,Bonilla:2014xla,Bonilla:2017ekt,Reig:2018ocz} and constitutes a key prediction of our model,
rather robust against renormalization group evolution. \\[-.4cm] 

Given the precise measurements of the masses of charged lepton $m_e$, $m_{\mu}$, $m_{\tau}$ and the bottom quark $m_b$, equation (\ref{eq:golden}) can be interpreted as a prediction for the
light quark masses $m_d$ and $m_s$. These are determined by lattice simulations with larger uncertainties~\cite{ParticleDataGroup:2022pth,FlavourLatticeAveragingGroupFLAG:2021npn}. 
As shown in Fig.~\ref{fig:Golden} it is in very good agreement with the reported values for the light quark masses at the $M_Z$ scale~\cite{Antusch:2013jca,Huang:2020hdv}.
One sees that not only the prediction is very close to the reported PDG values of the light quark masses,
but also the current central value (black star) got closer to the model's prediction than before (black dot).

\begin{figure}[t]
\centering
\includegraphics[height=6.5cm]{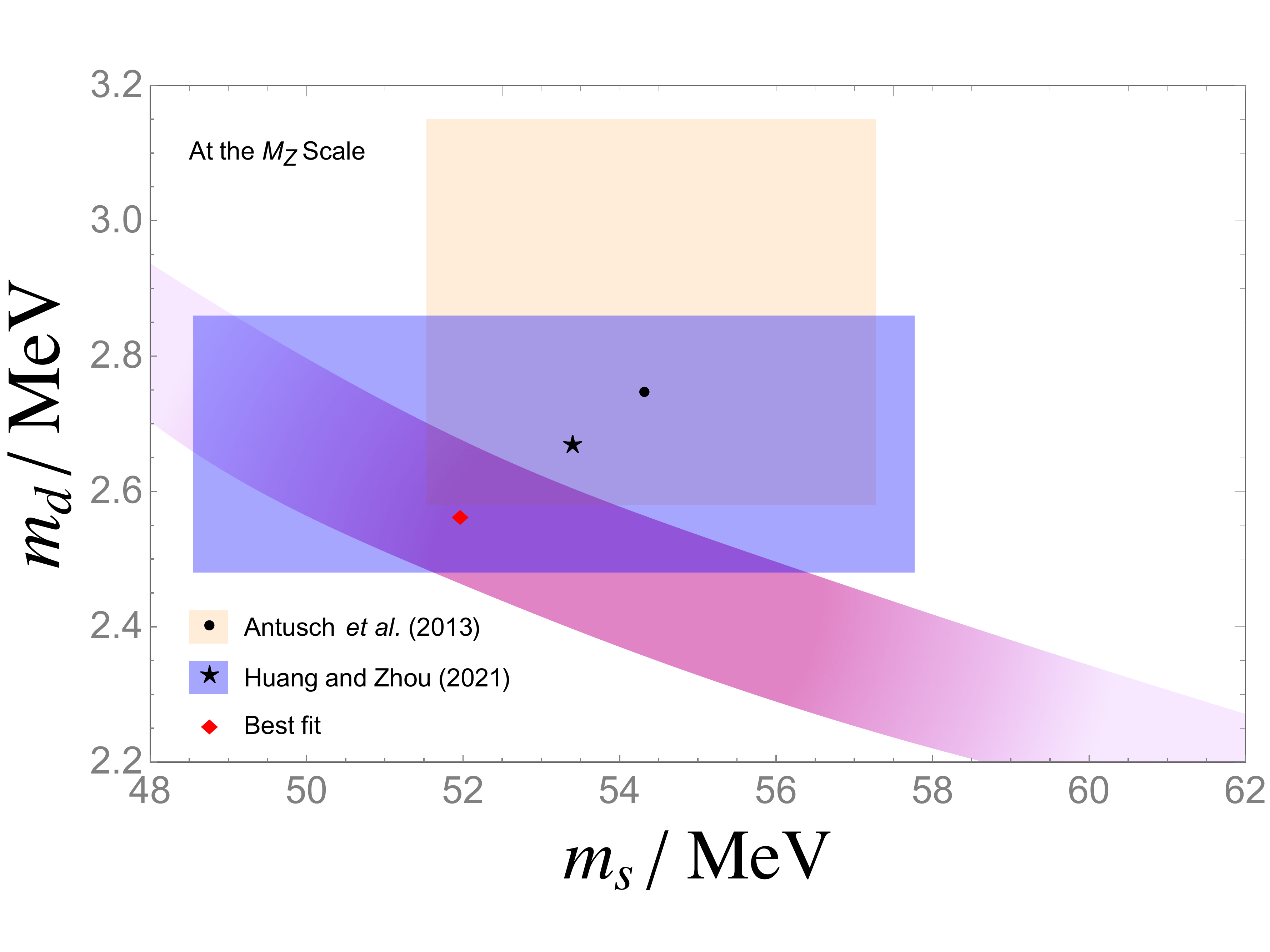}
\caption{ In magenta we give the 3$\sigma$-predicted light quark mass $m_d$ versus $m_s$ at the $M_Z$ scale, Eq.~(\ref{eq:golden}).
  In orange we display the 1$\sigma$ regions in each mass from~\cite{Antusch:2013jca}, and in blue the recent update in~\cite{Huang:2020hdv}.
  The model's best fit is denoted with a red diamond.}
\label{fig:Golden}
\end{figure}

\subsection{Neutrinoless Double Beta Decay} 
\label{sec:neutr-double-beta}

There are other relevant observables, such as the absolute neutrino mass scale, whose value is only bounded from above by current experiments.
Apart from cosmological observations~\cite{Planck:2018vyg}, we have constraints from tritium beta decay endpoint measurements at Katrin~\cite{KATRIN:2021uub},
as well as searches for neutrinoless double beta decay~\cite{KamLAND-Zen:2022tow}. 
The latter is a lepton number violating process and hence crucially depends on two additional physical phases~\cite{Schechter:1980gr,Schechter:1980gk} which are not included in the above discussion.
%
These phases are present in the lepton mixing matrix but do not manifest in neutrino oscillation experiments~\cite{Schechter:1980gk}.

The most convenient description of the lepton mixing matrix is the symmetrical parametrization proposed in~\cite{Schechter:1980gr} and revisited in~\cite{Rodejohann:2011vc}.  
The three physical phases are parametrized as $\phi_{12}$, $\phi_{13}$, and $\phi_{23}$, so that the leptonic ``Dirac'' $CP$ phase entering oscillations is given by
$$\delta^{\ell} = \phi_{13} - \phi_{12} -\phi_{23}$$
while the extra two phases are crucial to describe lepton number violating processes. 

In our model neutrinos are Majorana particles, so lepton number violating processes such as \znbb decay are expected. The associated amplitude is proportional to 
\begin{equation}
\vev{m_{\beta \beta}} = \left\lvert \cos^2\theta^{\ell}_{12} \cos^2\theta^{\ell}_{13} m^{\nu}_1 + \sin^2\theta^{\ell}_{12} \cos^2\theta^{\ell}_{13} m^{\nu}_2 e^{2 i \phi_{12}} +  \sin^2\theta^{\ell}_{13}  m^{\nu}_3 e^{2 i \phi_{13}}\right\rvert,
\end{equation}
Note that, as expected, $\vev{m_{\beta \beta}}$ only depends on the two Majorana phases, but not in the ``Dirac'' phase $\delta^{\ell}$~\footnote{In contrast to the PDG phase convention, the symmetrical parametrization
  provides a transparent description of \znbb decay.}.
 In Figure \ref{fig:Nuless-NO} we show the regions for the mass parameter $\vev{m_{\beta \beta}}$ characterizing the neutrinoless double beta decay amplitude, $\vev{m_{\beta \beta}}$. 
 The blue region is the generally allowed one, given the current experimental determination of neutrino oscillation parameters~\cite{deSalas:2020pgw},  
 while the magenta region gives the predicted region within our model, in which all of the 19 flavor observables in both the quark and lepton sectors, Eq.(\ref{eq:obs}),
 lie within 3-$\sigma$ of their measured values. 
  The best fit value is indicated as a red diamond, and lies somewhat below the present limit set by Kamland-Zen $(36 - 156\; \mathrm{meV})$ 
 which is shown as the upper orange horizontal band in Fig.~\ref{fig:Nuless-NO}.
 We have also displayed the projected sensitivities for some of the next generation experiments searching for \znbb, such as LEGEND \cite{LEGEND:2017cdu},
 SNO + Phase II \cite{Andringa:2015tza}, and nEXO \cite{Albert:2017hjq}, shown as horizontal dashed lines. 
 \begin{figure}[h]
\centering
\includegraphics[height=8cm]{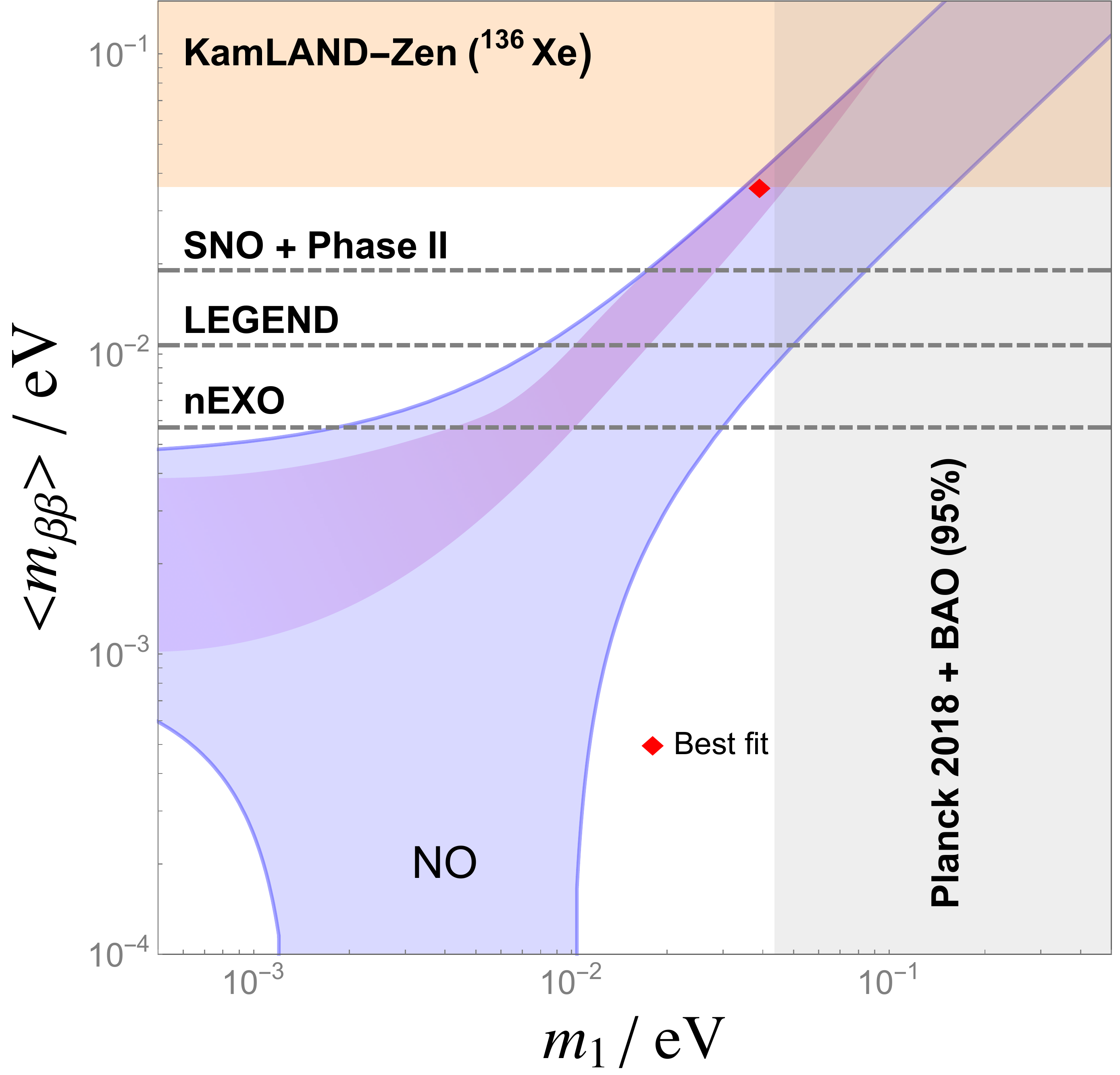}
\caption{ 
  Neutrinoless double beta decay mass parameter $\vev{ m_{\beta \beta}}$ versus the mass of the lightest neutrino $m_1$ for the normal ordered (NO) scenario.
  The blue region is allowed given the current neutrino oscillation parameters~\cite{deSalas:2020pgw}, while the predicted magenta region is the  one consistent with the masses and mixings
  of quarks and leptons at 3$\sigma$. The best fit point is indicated by a red diamond. The current bound on $\vev{ m_{\beta \beta}}$ from Kamland-Zen~\cite{KamLAND-Zen:2022tow} as well as
  future projections are shown as the upper horizontal band, and horizontal dashed lines respectively. The vertical band indicates the cosmological bound~\cite{Planck:2018vyg}.}
\label{fig:Nuless-NO}
\end{figure}
\begin{table}[ht]
	\centering
	\footnotesize
	\renewcommand{\arraystretch}{1.1}
	\begin{tabular}[t]{|lcr|}
		\hline
		Parameter &\qquad& Value \\ 
		\hline
		$y^e_1v_d/\mathrm{GeV}$ &\quad& $1.746$ \\
		$y^e_2v_d/(10^{-1}\mathrm{GeV})$ && $-9.78$ \\
		$y^d_1v_d/\mathrm{GeV}$ &&$-2.85$ \\
		$y^d_2v_d /(10^{-2}\mathrm{GeV})$ && $-7.35$ \\
		\rule{0pt}{3ex}%
		$(y_{1}^{\nu}v_\nu)^2 y^S v_{\sigma}/(m^2_{\nu^c S} $ $\mathrm{eV})$ &&$2.58$ \\
		$(y_{2}^{\nu}v_\nu)^2 y^S v_{\sigma}/(m^2_{\nu^c S} $ $\mathrm{eV})$ && $-0.54$ \\
		\rule{0pt}{3ex}%
		$y^u_{1} v_u/(10^{-3}\mathrm{GeV})$ &&$-4.765$ \\
		$y^u_2v_u/(10^{-1}\mathrm{GeV})$ && $3.582$ \\
		$y^u_3v_u/\mathrm{GeV}$ && $-7.08 $ \\
		\rule{0pt}{3ex}%
		$\epsilon^u_1/10^{-2}$ && $-8.65 $ \\
		$\epsilon^u_2 $ && $23.7$ \\
		\rule{0pt}{3ex}%
		$\epsilon^d_1/10^{-2}$ && $1.80$ \\
		$\epsilon^d_2/10^{-4}$ && $9.12$ \\
		\rule{0pt}{3ex}%
		$\epsilon^\nu_1$ && $1.749$ \\
		$\epsilon^\nu_2$ && $-7.397$ \\
		\rule{0pt}{3ex}%
		$(\phi^d_1-\phi^d_2)/\pi$ && $-0.076$ \\
		$(\phi^\nu_1-\phi^\nu_2)/\pi$ && $-0.321$ \\
		\hline	
	\end{tabular}
	\hspace*{0.5cm}
	\begin{tabular}[t]{ |l |c|c c |c| }
		\hline
		\multirow{2}{*}{Observable}& \multicolumn{2}{c}{Data} & & \multirow{2}{*}{Model best fit}  \\
		\cline{2-4}
		& Central value & 1$\sigma$ range  &   & \\
		\hline
		$\sin^2\theta_{12}^\ell$$/10^{-1}$ & 3.18 & 3.02 $\to$ 3.34 && $3.18$  \\ 
		$\sin^2 \theta_{13}^\ell$$/10^{-2}$ (NO) & 2.200 & 2.138 $\to$ 2.269  && $2.199$  \\  
		$\sin^2 \theta_{23}^\ell$$/10^{-1}$ (NO) & 5.74 & 5.60 $\to$ 5.88  && $5.74$ \\ 
		$\delta^\ell$$/ \pi$ (NO) & 1.08 & 0.96 $\to$ 1.21 && $1.12$  \\
		$m_e$$/ \mathrm{MeV}$ & 0.486 &  0.486 $\to$ 0.486 && $0.486$ \\ 
		$m_\mu$$/  \mathrm{GeV}$ & 0.102 & 0.102  $\to$ 0.102  &&  $0.102$ \\ 
		$m_\tau$$/ \mathrm{GeV}$ &1.746 & 1.746 $\to$1.746 && $1.746$ \\ 
		$\Delta m_{21}^2 / (10^{-5} \, \mathrm{eV}^2 ) $ & 7.50  & 7.30 $\to$ 7.72 && $7.50$  \\
		$\Delta m_{31}^2 / (10^{-3} \, \mathrm{eV}^2) $ & 2.55  & 2.25 $\to$ 2.75 &&  $2.55$ \\
		$m_1$$/\mathrm{meV}$  & & & & $39.02$ \\ 
		$m_2$$/\mathrm{meV}$  & && & $39.97$ \\ 
		$m_3$$/\mathrm{meV}$  & && & $ 63.81$ \\
		$ \phi_{12} $  & & && $0.245$  \\
		$ \phi_{13}$  & & && $5.22$  \\    
		$ \phi_{23}$  & & &&  $1.46$ \\   
	       $ \langle m_{\beta \beta} \rangle$$/ \mathrm{eV}$ & & &&  $0.036$ \\
		\hline
		$\theta_{12}^q$ $/^\circ$ &13.04 & 12.99 $\to$ 13.09 &&  $13.03$ \\	
		$\theta_{13}^q$ $/^\circ$ &0.20 & 0.19 $\to$ 0.22 && $0.21$  \\
		$\theta_{23}^q$ $/^\circ$ &2.38& 2.32 $\to$ 2.44 && $2.41$  \\	
		$\delta^q$ $/^\circ$ & 68.75 & 64.25 $\to$ 73.25  & & $69.11$\\
		$m_u$ $/ \mathrm{MeV}$ & 1.23 & 1.08$\to$ 1.51 && $1.23$  \\	
		$m_c$ $/ \mathrm{GeV}$ & 0.620 & 0.603 $\to$ 0.637 &&  $0.620$ \\	
		$m_t$ $/\mathrm{GeV}$  	  & 168 & 167 $\to$ 169 && $168$ \\
		$m_d$ $/ \mathrm{MeV}$ & 2.67& 2.57 $\to$ 2.94 &&  $2.57$ \\	
		$m_s$ $/ \mathrm{MeV}$ & 53.1 & 51.61 $\to$ 58.32 && $51.7$ \\
		$m_b$ $/ \mathrm{GeV}$	  & 2.84 & 2.76 $\to$ 2.87 &&  $2.81$\\
		\hline
		$\chi^2_K$ & & & & $1.05$ \\
		\hline
		
	\end{tabular}
        \caption{
          Summary of our goodness-of-fit global analysis of flavor observables.
          Charged lepton masses are taken from~\cite{Antusch:2013jca} and quark masses are based on~\cite{Huang:2020hdv}.
          Neutrino oscillation parameters are taken from~\cite{deSalas:2020pgw} assuming normal-ordering. }  
	\label{tab:fit2}
\end{table}

Let us now comment on Table~\ref{tab:fit2}.
As already noted, the best fit point gives $\chi^{2}_{K} = \frac{\chi^2}{K}=1.05$ showing that indeed our model is in excellent agreement with experiment.
Note also that one of the phases in the fit, $\phi^d_1-\phi^d_2$, practically vanishes. This would suggest that a reasonable fit of the flavor parameters could be achieved even with these phase
parameters fixed to vanish. Without them, one would still find a not-so-bad description, with a reduced chi-squared $\chi^2_K=5.8$.
The main tension with experiment for this case lies in the physical $CP$ phases $\delta^l,\delta^q$, of about $3\sigma$.
This would be the price to pay for having an enhanced predictivity for other flavour observables.
The vanishing of these phases could follow from the imposition of a generalized CP symmetry, in the manner described in \cite{deAnda:2019jxw}.

\section{Discussion and outlook }
\label{sec:discussion-outlook-}

In this paper we have examined a low-scale realization of the orbifold theory of flavor proposed in~\cite{deAnda:2019jxw,deAnda:2020pti}. 
Below compactification an $A_4$ family symmetry emerges naturally, providing a good description of flavor physics, see Table~\ref{tab:fit2}.  
The first Kaluza-Klein excitation of the same scalar which drives family symmetry breaking as well as neutrino mass generation through the inverse seesaw mechanism
can be identified with WIMP Dark matter, see Fig.~\ref{fig:DMplot}. 
The model predicts the ``golden'' quark-lepton mass relation, Eq.~(\ref{eq:golden}) and Fig.~\ref{fig:Golden}.
Concerning neutrinos we have interesting \znbb decay predictions shown in Fig.~\ref{fig:Nuless-NO}.
For definiteness we have assumed the favored case of normal neutrino mass ordering.
Let us also mention that our model is not necessarily meant to be a UV-complete construction. 
For instance, the fact that there are three color triplets in the bulk generates 6-dimensional gauge anomalies.
However, after compactification, at low energies, these anomalies cancel and the theory is consistent.
One may assume that there are extra fields at high energies that cancel the anomalies at that level too~\cite{Dobrescu:2001ae}.

All in all, our model provides an excellent global description of flavor observables both in the quark as well as lepton sectors starting from first principles.
Compared with the work in Refs.~\cite{deAnda:2019jxw,deAnda:2020pti} the results given here correspond to a low-scale inverse seesaw mechanism
 and they follow from the use of updated flavor parameters, such as oscillation parameters as well as recent quark mass determinations.

\acknowledgements 
\noindent

Work supported by the Spanish grants PID2020-113775GB-I00 (AEI/10.13039/501100011033) and Prometeo CIPROM/2021/054 (Generalitat Valenciana).
OM is supported by  Programa Santiago Grisolía (No. GRISOLIA/2020/025).
CAV-A is supported by the Mexican C\'atedras CONACYT project 749 and SNI 58928. The relic abundance and direct detection constraints were
calculated using the micrOMEGAS package \cite{Belanger:2018ccd} at GuaCAL (Guanajuato Computational Astroparticle Lab). We thank Ignatios Antoniadis for discussions in the early phase of the work.


\bibliographystyle{utphys}
\bibliography{bibliography}
\end{document}